\documentclass[11pt,a4paper,oneside,thmsb,notitlepage]{article}
\usepackage{amsfonts}
\usepackage{amssymb}
\usepackage{latexsym}
\usepackage{enumerate}
\usepackage{amsmath}
\usepackage{amsthm}
\usepackage{graphicx}
\usepackage{rotating}
\usepackage{multirow}
\usepackage{bm}
\usepackage{caption}
\usepackage{subcaption}
\usepackage{graphicx}

\renewcommand{\baselinestretch}{1.5}

\begin{document}

\title{The sooner the better: lives saved by the lockdown during the COVID-19 outbreak. The case of Italy}
\author{Roy Cerqueti$^{a,b}$, Raffaella Coppier$^{c}$, Alessandro Girardi$^{d}$\thanks{Any opinion, finding, and conclusion or recommendation expressed in this material are those of the author(s) and do not necessarily reflect the views of the PBO.}, Marco Ventura$^{e}$\thanks{%
Corresponding author: Marco Ventura, Department of Economics and Law, Sapienza University of Rome, Via del Castro Laurenziano 9- 00161, Rome, Italy. Fax: +39 (0)6 233 232 419. Email: marco.ventura@uniroma1.it } \\
{\small $^a$ Department of Social and Economic Sciences -- Sapienza University of Rome, Italy}\\
{\small $^b$ School of Business -- London South Bank University, UK}\\
{\small Email: roy.cerqueti@uniroma1.it}\\
{\small $^c$ Department of Law and Economics -- University of Macerata} \\
{\small Email: raffaella.coppier@unimc.it}\\
{\small $^d$ Parliamentary Budget Office, PBO, Rome, Italy} \\
{\small Email: alessandro.girardi@upbilancio.it} \\
{\small $^{e}$  Department of Economics and Law -- Sapienza University of Rome, Italy}\\
{\small Email: marco.ventura@uniroma1.it} } \maketitle

\begin{abstract}
This paper estimates the effects of non-pharmaceutical interventions – mainly, the lockdown – on the COVID-19 mortality rate for the case of Italy, the first Western country to impose a national shelter-in-place order. We use a new estimator, the Augmented Synthetic Control Method (ASCM), that overcomes some limits of the standard Synthetic Control Method (SCM). The results are twofold. From a methodological point of view, the ASCM outperforms the SCM in that the latter cannot select a valid donor set, assigning all the weights to only one country (Spain) while placing zero weights to all the remaining. From an empirical point of view, we find strong evidence of the effectiveness of non-pharmaceutical interventions in avoiding losses of human lives in Italy: conservative estimates indicate that for each human life actually lost, in the absence of lockdown there would have been on average other 1.15, the policy saved in total 20,400 human lives.
\end{abstract}

\renewcommand{\baselinestretch}{1}


\section{Introduction}
\label{sec:intro}

Exponentially growing threats require strong and early policy response. In the first wave of the COVID-19 outbreak, the timing of confinement measures played a fundamental role in flattening the contagion curve (Flaxman et al., 2020; Amuedo-Dorantes et al., 2020). However, policy makers reasonably hesitate to take resolute measures when threats appear to be limited. This caution is reasonable, because if countermeasures work, it will seem in retrospect as if the policy response was an overreaction, possibly causing a loss of consensus (Pisano et al., 2020; The Economics, 2020). Once the curve is flattened, the public will likely blame the incumbent government for the tremendous economic losses caused by social distancing orders without fully grasping their essential role in halting the spread of the viral disease.
\newline
\indent This paper adds to this debate, estimating the effects of lockdown -- or, in general, of the so-called non-pharmaceutical interventions – on the propagation of COVID-19 with particular emphasis on the most relevant aspect: saving human lives. Its real-world relevance, implications and its high level of socio-economic meaningfulness need no further explanation. The task is particularly challenging from a methodological perspective due to typical selection bias problems, and this explains the growing interest of econometricians in this research question. At first instance, a natural candidate to face this challenge is the SCM – first introduced by Abadie and Gardeazabal (2003) and the subsequent studies by Abadie et al. (2010, 2015).\footnote{Since its introduction, the SCM has been widely used in social sciences and applied to a broad spectrum of topics, spanning from terrorism and crime to natural resources and disasters, political and economic reforms, immigration, education, pregnancy and parental leave, taxation, as well as social connections and local development. Athey and Imbens (2017) have defined it as the most crucial innovation in the policy evaluation literature over the last fifteen years. For a recent survey see Abadie (2020).} At its very essence, the SCM involves the comparison of outcome variables between the treated unit, i.e., the unit affected by the intervention and similar but different unaffected units, reproducing an accurate counterfactual of the unit of interest in the absence of intervention, commonly referred to as the synthetic unit. This method can be conceived as a data-driven procedure to retrieve non-treated units sharing similar characteristics concerning the treated in the pre-intervention period. In the COVID-19 context, the treatment consists of non-pharmaceutical interventions implemented at a given point in time in a specific country or, in general, in a region, so that the SCM lends itself as one of the most immediate tools to face the problem empirically. For this reason, it has been (more or less successfully) repeatedly applied in the very recent literature on COVID-19. Bayat et al. (2020) and Friedson et al. (2020) have applied the SCM on US data, where the latter takes a specific focus on California, taking advantage of the timely enactment of the shelter-in-place order. Friedson et al. (2020) extend the analysis of the impact to the unavoidable economic drawbacks, such as the number of jobs lost. As far as the measurement of the avoided virus spread is concerned, it is possible to mention Huber and Langen (2020), who apply the SCM design to Switzerland, and Tian et al. (2020a, 2020b) to China. Some authors have estimated the effect of particular prescriptions included in the more general lockdown policy. Notably, Mitze et al. (2020) study the effect of wearing face masks in Germany and Neidh\"{o}fer and Neidh\"{o}fer (2020) study the effect of school closures for Argentina, Italy and South Korea. Lee and Yang (2020) offer a purely economic perspective focusing on the impact of contrasting COVID-19 measures on the labour market in South Korea. As a peculiar case, Sweden is a country that has attracted a lot of attention in this growing literature because it is one of the very few countries that has not enacted a statewide shelter-in-place order (to date). Due to its unusual situation, Sweden offers the researcher the opportunity to look at the case as if the treatment consisted in non-intervening \textit{vis-a-vis} a group of countries in which interventions took place (see Born et al., 2020 and Cho, 2020).  \\
Despite its popularity, however, some real-world circumstances can make this instrument unapplicable, under penalty of biased estimates. This occurrence may be due to many reasons, such as the employment of different heterogeneous measurement methodologies whereby the same phenomenon is measured in different countries. The case under scrutiny falls exactly within these circumstances. To overcome this problem, we have followed a general approach proposed by Ben-Michael et al. (2020), augmenting the SCM with the Ridge regression model, therefore obtaining a Ridge ASCM, which can be written as a weighted average of the control unit outcomes. To the best of our knowledge, the Ridge ASCM has not yet been applied to assess the impact of non-pharmaceutical interventions for the COVID-19 pandemic. It is possible to find just a couple of papers in the environmental field (notably, forest fires in Colombia, Amador-Jimenez et al., 2020, and air pollution in China, Cole et al., 2020).

Our contribution to the literature on the socio-economic effects related to the COVID-19 pandemic is manifold. First, it sheds light on a controversial policy intervention that has been and still is largely debated. The intervention's effectiveness is evaluated in terms of the most immediate and desired effect, namely avoided deaths. Secondly, from a strictly methodological point of view, it uses one of the most recent advances of the popular SCM, i.e., the Ridge ASCM, which allows one to overcome the non-negligible limit of non-perfect pre-treatment fit, which, in turn, would generate a biased estimate from the canonical estimator.\footnote{Other recent advances pertain to multiple treated units (Robbins et al., 2017; Abadie and L'Hour, 2019), extensions of permutation methods (Dube and Zipperer, 2015), denoising the outcome variable and imputation of missing values (Amjad et al. 2018, 2019 and Athey et al. 2018), inference (Chernozhukov et al. 2019a, 2019b; Cattaneo et al. 2019), the role of covariances (Botosaru and Ferman, 2019; Ferman et al., 2020) and bias correction (Powel, 2018; Arkhangelesky et al., 2019; Chernozhukov et al., 2019b).} Thirdly, the study focuses on Italy, which is widely acknowledged as a paradigmatic case, owing to its pioneering role -- immediately after China -- in facing the current pandemic disease. The first Italian COVID-19 cases were registered quite early (January 2020 or even before) and contagion has accelerated since its inception. In March 2020, Italy was the country with the highest number of cases -- apart from China -- rapidly becoming the European epicentre of outbreak, with 207,428 confirmed cases and 28.236 deaths as of the beginning of May 2020 (Ministry of Health). These figures represented approximately 14\% of all confirmed cases and 20\% of deaths in Europe, 6\% of confirmed cases and just over 11\% of deaths worldwide. Moreover, and even more interestingly from our standpoint, Italy was the first Western country in which the government imposed restrictions on mobility, economic activities and social interactions -- the already mentioned strict lockdown. The lockdown order was officially imposed in Italy from March $9$ up to the May $18$, 2020, 70 days. The intervention was highly criticized at that time, possibly because it was the first among Western countries, and it was not yet completely clear the importance of acting timely, especially on the part of some media and politicians. Nonetheless, many other European countries followed the Italian model within a few weeks, including the UK that initially claimed to be against such a type of intervention. A long list of scientific contributions witnesses the relevance of the Italian case for understanding the COVID-19 spread and, consequently, the effectiveness of non-pharmaceutical interventions (see, e.g., Bonacini et al., 2020; Cameletti, 2020; Eckardt et al., 2020; Lolli et al., 2020; Palladino et al., 2020, Peracchi and Terlizzese, 2020). In particular, the works by Modi et al. (2020) and Cerqua et al. (2020) deserve special mentioning because both use the standard SCM to assess the plausibility of official figures, concluding that the true count of deaths due to COVID-19 is supposed to be significantly higher than the official count. Given the very high relevance of the issue at stake, we contribute to the debate over the effects of lockdown measures by using the Ridge ASCM and enjoying all its methodological advances. To the best of our knowledge, this is the first paper dealing with applying the Ridge ASCM to Italy's paradigmatic case and evaluating the effects of a lockdown in terms of public health. Operatively, the task mentioned above is particularly challenging for at least a couple of other reasons. First, all the other European countries (the most similar to the one under scrutiny) were sooner or later treated. Secondly, the virus spread followed different paths among countries at different points in time. Both features make constructing a credible counterfactual series particularly difficult, and this pitfall is exacerbated by the limits the standard SCM suffers from. \\
Our results show that the SCM cannot generate a valid counterfactual, collapsing all the weights on one single unit, i.e., Spain, so that the lockdown effect would be calculated as a naive difference between the two countries. Allowing for negative weights, we provide evidence about the effectiveness of non-pharmaceutical interventions in saving human lives and avoiding a collapse of the Italian health care system. The take-home message from this study is twofold. From a methodological point of view, it suggests that the researcher must compare the weights before choosing whether to apply the SCM or the ASCM. A substantial distance of the two may be an indirect indicator of bias in the SCM estimates. From a socio-economic perspective, it documents that non-pharmaceutical intervention has led to at least 1.15 people saved for each life lost, for a total of 20,400 lives within the first 35 days of lockdown. \\
The remainder of the paper is organized as follows. Section \ref{sec:2} introduces the Ridge ASCM as an extension of the canonical SCM and describes the estimation exercise dataset. Section \ref{sec:3} presents the empirical set up paying particular attention to how the benchmark specification is chosen and presents the estimates under testing in Section \ref{sec:sensitivity}. Finally, Section \ref{sec:conclusion} proposes some back-of-the-envelope computations based on the estimates and draws some conclusions.

\section{Methodology and Data}
\label{sec:2}
This section is devoted to the illustration of the Ridge ASCM procedure. To be self-contained and provide a better understanding of this econometric method, we begin with a brief description of the canonical SCM. In its very essence, the SCM aims to simulate the outcome path of a country if it did not undergo a particular policy intervention. Operatively, the synthetic control is built as a weighted average of the units in the control group (donor pool), where the weights are chosen so that the synthetic control's outcome closely matches the treated unit's trajectory in the pre-treatment period, while also satisfying some constraints such as being non-negative or adding up to one.

\subsection{Econometric framework}
\label{SCM}
More formally, let $Y_{it}(0)$ and $Y_{it}(1)$ represent the potential outcomes for unit $i$, with $i=1, \dots, N$, at time $t$, with $t=1, \dots, T$, under control and treatment, respectively. Let $W_i$ be an indicator that unit $i$ is treated at time $T_0<T$, where units with $W_i=0$ never receive the treatment. In the SCM, one supposes that only one unit receives treatment and for ease of reference this is listed as the first one, $W_1=1$, the remaining $N_0=N-1$ units are possible controls, usually referred to as donor units. Without loss of generality, for the time being, the post-treatment observation is limited to just one: $T=T_0+1$. The observed outcomes are then

\begin{equation}
\label{POM}   Y_{it}=\left\{
	\begin{array}{l}
				Y_{it}(0) \quad \; if \; W_i=0 \; or \; \; \; t\leq T_0  \\
				Y_{it}(1) \quad \; if \; W_i=1 \; and \; t> T_0
					\end{array}
\right.
\end{equation}
We also assume that control potential outcomes are generated as a fixed component $m_{it}$ plus a mean-zero additive noise $\varepsilon_{it}$ drawn from some distribution,
\begin{equation}
Y_{it}(0)=m_{it}+\varepsilon_{it}
\end{equation}
The treated potential outcome is then $Y_{it}(1)=Y_{it}(0)+\tau_{it}$ where $\tau_{it}$ represent the treatment effects -- which are the objects of our estimation -- and are fixed parameters. Therefore, the treatment effects can be rewritten as $\tau=\tau_{1T}=Y_{iT}(1)-Y_{iT}(0)$. The error terms in the post-treatment period are collected in the vector $\bm{\varepsilon}_{T}=(\varepsilon_{1T}, \dots, \varepsilon_{NT})$ and are assumed to be mean-zero and uncorrelated with treatment assignment. That is, the treatment assignment $W_i$ is ignorable given $m_{it}$,
\begin{equation}
\mathbb{E}_{\varepsilon_T}\left[W_i \varepsilon_{iT}\right]=\mathbb{E}_{\varepsilon_T}\left[\left(1-W_i \right)\varepsilon_{iT}\right]=\mathbb{E}_{\varepsilon_T}\left[ \varepsilon_{iT}\right]
\end{equation}
where $\mathbb{E}_{\varepsilon_T}$ denotes the expectation taken with respect to the error term $\bm{\varepsilon}_{T}$. It follows that the noise terms for the treated and control units do not systematically deviate from each other. Let $X_{it}$ represent pre-treatment outcomes that are used as and along with other covariates, $\mathbf{X_0}$ represents the $N_0 \times T_0$ matrix of control units pre-treatment outcome and covariates. $\mathbf{Y}_{0T}$ is the $N_0$ vector of control unit outcomes in period $T$. With only one treated unit, $Y_{1T}$ is a scalar, and $X_1$ is a $T_0$-row vector of treated unit pre-treatment outcomes and/or covariates. The potential outcome for the treated unit, $Y_{1T}(0)$, is computed by the SCM as a weighted average of the control outcomes, $\mathbf{Y}_{0T}^\prime \mathbf{\gamma}$, being $\bm{\gamma}=(\gamma_1, \dots, \gamma_N)$ the vector of weights. The elements of $\bm{\gamma}$ are chosen to balance pre-treatment outcomes and other covariates.\footnote{Notice that since pre-treatment outcomes are included in the $\mathbf{X_0}$ matrix, ``pre-treatment fit'' and ``covariance balancing'' are equivalent expressions.} To our aim, the SCM can be formalized as a solution with respect to $\bm{\gamma}$ of the following constrained optimization problem

\begin{equation}
\label{SCM1}
	\begin{array}{l}
		{\rm min}_{\gamma} \quad \left\|\left( \bm{X_1-X_{0}^{\prime}\gamma} \right)  \right\|_{2}^{2}+\zeta \sum_{W_i=0} f(\gamma_i) \\
		s.t. \qquad	\sum_{W_i=0}\gamma_i=1 \\
		\qquad \qquad \gamma_i\geq0 \; \quad i:W_i=0 \\
\end{array}
\end{equation}

\noindent where the constrains limit $\bm{\gamma}$ to the unit simplex and where \\ $\left\|\left( \bm{X_1-X_{0}^{\prime}\gamma} \right)  \right\|_{2}^{2} \equiv \left( \bm{X_1-X_{0}^{\prime}\gamma} \right)^{\prime}\left( \bm{X_1-X_{0}^{\prime}\gamma} \right)$ is the 2-norm on $\mathbb{R}^{T_0}$. The simplex constraint in (\ref{SCM1}) ensures that the weights will be sparse and non-negative, while the hyperparameter $\zeta>0$ penalizes the dispersion of the weights, following a suggestion by Abadie et al. (2015). The optimization problem in (\ref{SCM1}) can be regarded as an approximate balancing weights estimator. These weights achieve perfect pre-treatment fit, and the resulting estimator has many attractive properties including a bias bound derived by Abadie et al. (2010), when the treated unit's vector of lagged outcomes and covariates, $\mathbf{X_1}$, is inside the convex hull of the control units' lagged outcomes and covariates, $\mathbf{X_0}$. Due to possible high dimension, however, achieving perfect pre-treatment fit is not always feasible with weights constrained to be on the simplex and in these cases Abadie et al. (2015) recommend against using SCM. Thus, the conditional nature of the analysis is critical to deploying SCM, excluding many practical settings. This drawback is overcome by the Ridge ASCM of Ben-Michael et al. (2020), which is now discussed. \\
The Ridge ASCM proposes modifying the problem in (\ref{SCM1}) as follows:
\begin{equation}
\label{ASCM1}
\hat{Y}_{1T}^\prime(0)=\sum_{W_i=0}\hat{\gamma}_i^{scm}Y_{iT}+\left(\hat{m}_{1T}-\sum_{W_i=0}\hat{\gamma}_i^{scm}\hat{m}_{iT} \right)
\end{equation}

\begin{equation}
\label{ASCM2}
=\hat{m}_{1T}+\sum_{W_i=0}\hat{\gamma}_i^{scm}\left(Y_{iT}- \hat{m}_{iT}\right)
\end{equation}
where $\hat{\gamma}_i^{scm}$ is the estimated $i$-th SCM weight. In this context, the canonical SCM is a special case in which $\hat{m}_{iT}$ is constant.
Albeit fully equivalent, equations (\ref{ASCM1}) and (\ref{ASCM2}) highlight two distinct features of the Ridge ASCM. From (\ref{ASCM1}) it appears clear that Ridge ASCM corrects the SCM estimate, $\sum_{W_i=0}\hat{\gamma}_i^{scm}Y_{iT}$, by the imbalance in a particular function of the pre-treatment, $\hat{m}(.)$. Intuitively, since $\hat{m}(.)$ estimates the post-treatment outcome, we
can view this as an estimate of the bias due to imbalance, analogous to bias correction for inexact matching (Rubin, 1973; Abadie and Imbens, 2011). Therefore, the SCM and Ridge ASCM estimates will be similar if the estimated bias is small. 
Differently, equation (\ref{ASCM2}) is similar in spirit to standard doubly robust estimation (Robins et al.,
1994), which begins with the outcome model but then re-weights to balance residuals. Given these premises, the choice of the  estimator $\hat{m}(.)$ is important both to understand the properties of the procedure and for practical performance. Notably, there are some attractive features of estimating  $\hat{m}(.)$ via Ridge regression which is linear in both pre-treatment outcomes and in comparison units. This case is referred to as the Ridge ASCM. In this case, the estimator of the post-treatment outcome is $\hat{m}(\mathbf{X}_i)=\hat{\eta}_0^{ridge}+ \mathbf{X_i^\prime \hat{\bm{\eta}}}^{ridge}$, where $\hat{\eta}_0^{ridge}$ and  $\bm{\hat{\eta}}^{ridge}$ are the coefficients of a Ridge regression of control post-treatment outcomes $Y_{0T}$ on centered pre-treatment outcomes $\mathbf{X}_0$ with penalty hyperparameter $\lambda^{ridge}$:

\begin{equation}
 \left\{\hat{\eta}_0^{ridge},\bm{\hat{\eta}}^{ridge} \right\} = {\rm argmin}_{\eta_0,  \bm{\eta}} \frac{1}{2} \sum_{W_i=0} \left( Y_i- \left(\eta_0+X_i^{\prime} \bm{\eta} \right) \right)^{2} + \lambda^{ridge} \left\| \bm{\eta}  \right\|_{2}^{2}
\end{equation}

The Ridge ASCM estimator is then:
\begin{equation}
\hat{Y}_{1T}^{aug}(0) = \sum_{W_i=0} \hat{\gamma}_i^{scm} Y_{iT}+ \left(\mathbf{X}_1- \sum_{W_i=0} \hat{\gamma}_i^{scm} \mathbf{X}_{i} \right) \cdot \hat{\bm{\eta}}^{ridge}
\end{equation}
when augmenting with Ridge regression the implied weights are themselves the solution to a penalized synthetic control problem, as in the standard SCM problem. Nevertheless, while the original SCM constrains weights to be on the simplex, this does not occur with the Ridge ASCM. Indeed, when the treated unit lies outside the convex hull of the control units, the Ridge ASCM improves the pre-treatment fit relative to the SCM by allowing for negative weights and extrapolating away from the convex hull. The Ridge ASCM directly penalizes the distance from the sparse, non-negative SCM weights, controlling the amount of extrapolation by the choice of $\lambda^{ridge}$, and only resorts to negative weights if the treated unit is outside of the convex hull. When the treated unit is in the convex hull of the control units -- so the SCM weights exactly balance the lagged outcomes -- the Ridge ASCM and SCM weights are identical. When the SCM weights do not achieve an exact balance, the Ridge ASCM solution will use negative weights to extrapolate from the convex hull of the control units. The amount of extrapolation is determined both by the
amount of imbalance and by the hyperparameter $\lambda^{ridge}$. When SCM yields good pre-treatment fit or when $\lambda^{ridge}$ is large, the adjustment term will be small and $\gamma^{aug}$ will remain close to the SCM weights, Ridge ASCM and SCM weights will be equivalent and the estimation error will only be due to variance of the weights and post treatment noise. It follows that $\lambda^{ridge}$ plays a crucial role and its value must be derived optimally. Operatively, one possibility is to follow the in-time placebo check proposed by Abadie et al. (2015). Let $\hat{Y}_{1k}^{(-t)}=\sum_{W_i=0}\hat{\gamma}_{i(-t)}^{aug}Y_{ik}$ be the estimate of $Y_{1k}$ obtained excluding time period $t$ from the sample. The idea consists in comparing the difference $Y_{1t}-\hat{Y}_{1t}^{(-t)}$ for some $t \leq T_0$ as a placebo check. We can extend this idea to compute the leave-one-out cross validation Mean Squared Error (MSE) over time periods:
\begin{equation}
\label{CV}
CV(\lambda^{ridge})=\sum{}_{t=1}^{T_0} \left(Y_{1t}-\hat{Y}_{1t}^{(-t)} \right)^2
\end{equation}
and the cross validation procedure chooses either the lambda that minimizes (\ref{CV}) or the maximal value of $\lambda^{ridge}$ with MSE within one standard deviation of the minimal MSE, as suggested by Hastie et al. (2009), among others. Such a choice trades off overfitting, i.e. a too small $\lambda^{ridge}$, and biased estimates, i.e. a too large $\lambda^{ridge}$, indeed if $\lambda^{ridge}\rightarrow\infty$ ASCM is equivalent to SCM.

\subsection{Data sources and variables construction}
\label{sec:datasource}
As mentioned above, the ASCM\footnote{For ease of reference, henceforth the simpler expression ``ASCM'' will be used referring to ``Ridge ASCM''} procedure's goal is to evaluate the impact of a lockdown on the most immediate and desired effect, namely the number of avoided deaths.\footnote{Cho (2020) pointed out that identifying the most appropriate outcome variable to assess real epidemiological effects is controversial. On the one hand, endogenous cross-country differences in testing rates regarding eligibility and accessibility limit the usefulness of the cases of infection. On the other hand, daily death counts might be affected by measurement problems because some jurisdictions include both confirmed and probable cases of deaths, as opposed to others only reporting confirmed cases.} Specifically, the outcome variable $\textbf{Y}$ is the mortality rate, which is defined as the cumulative death counts per million population (\textit{dth}) taken from the Epidemic Intelligence team of the ECDC (European Center for Disease Prevention and Control). Since daily reported figures for deaths tend to be challenging to compare and qualify across countries due to possible confounding idiosyncratic socioeconomic differences related to health care systems and population ageing, we also consider several covariates, \textbf{X}, that are expected to be linked to the outcome variable. Accordingly, among the predictors, we include variables capturing the COVID-19 dynamics, such as cumulative cases per million population (\textit{num}), which are intuitive predictors of mortality rates. The second group of predictors includes variables capturing the ``resilience'' of each country's health system. This subset includes the number of hospital beds per hundred thousand population (\textit{hsp}) under the assumption that the more developed the health system, the less fatal the COVID-19 infection will be. As for the outcome variable, the source for \textit{num} and \textit{hsp} is the Epidemic Intelligence team of the ECDC. Following S\'a (2020) and Rockl\"ov and Sj\"odin (2020), among others, we also include socioeconomic characteristics that are likely to be (positively) related to mortality rates; accordingly, the median age (\textit{age}), as well as the average household size (\textit{hld}), are added to the set of covariates. All demographic variables are taken from the United Nations report (United Nations, 2019). We also control for ``mobility trends'' across different categories of places  and behavior changes derived from Google Mobility Reports, which collect percentage changes in visits and length of stay at different places relative to a baseline given by the median values of the same day of the week from January 3, 2020, to February 6, 2020.\footnote{This type of data is collected from smartphones with an initial level of ``normal conditions'' which is set to 0. When this data is reduced from the base value, it suggests that some forms of mobility constraints have been imposed in a specific area so that the average mobility decreases.} Following Chernozhukov et al. (2021), we focus on four out of six mobility sub-indices (namely, ``Grocery and Pharmacy'', ``Transit Stations'', ``Retail and Recreation'' and ``Workplaces''). ``Parks'' and ``Residential'' are dropped because the former does not have clear implications on the spread of COVID-19, while the latter shows an overly-high correlation with ``Workplaces'' and ``Retail and Recreation''. We distil the information content conveyed by the Mobility indicators into a synthetic index (\textit{mob}) by following a ``nonmodel based'' aggregation scheme, as discussed in Marcellino (2006).\footnote{Specifically, the four sub-indices are standardised to have zero mean and unit standard deviation. This step helps avoid the resulting (simple) average index, which is calculated in the subsequent step, to be dominated by variables with a particularly pronounced degree of volatility or an incomparably high absolute mean.} The subsequent logical step is identifying the donor states to form the synthetic control unit. When constructing a reliable counterfactual, it is well understood that the relationship between the predictors and the outcome variable in the donor pool must be as similar as possible to the relationship in the treated unit. Accordingly, the selection of the donor pool's candidate elements should be carried out by identifying countries sharing some key similarities to the treated one. In the present context, geographical proximity is a crucial factor to be considered as the spread of the pandemic has been not homogeneous across space and over time, moving from Asia in late 2019 to Europe at the beginning of 2020 and, subsequently, to the Americas. Given our focus on the Italian case, an obvious choice to select the donor pool's elements is to focus on European countries. Accordingly, we have included all members belonging to the European Union (except Luxembourg) plus Switzerland, Norway, and the United Kingdom (28 countries in total). \\
Since the daily evolution of the mortality rate at the individual country level reflects different diffusion patterns at a given point in time, we have normalized the time unit such that ``day 1'' refers to the day on which cumulative infection cases per million exceeds one in the treated country as in Cho (2020). In our case, ``day 1'' corresponds to February 23, 2020, with the lockdown policy enacted on March 9. Therefore, in our setup, the pre-treatment period consists of 15 daily observations. Because the treated state contrasts to the control unit after treatment, the relevant policy under scrutiny (the impact of non-pharmaceutical interventions in our context) should not be enacted in any donor pool state during the study. Accordingly, our sample's ending date is given by the date when lockdown measures have taken place in the synthetic counterfactual. To identify such an average date, we have used an ad hoc index elaborated by the Oxford COVID-19 Government Response Tracker, namely the Stringency Index, SI, which collects standardized information on several different common government responses daily for a large number of countries. More specifically, we have followed Cho (2020) and defined the ending sample as the date on which the SI peaked in each donor country. Therefore, our sample's last observation occurs on ``day 149', corresponding to April 12, 2020, so our post-treatment period consists of 34 data points.

\section{Empirical setup}
\label{sec:3}
\subsection{Specification searching}
\label{sec:specification}
Resorting to the ASCM makes it possible to choose in a transparent way the weights to build the counterfactual for the treated unit (Abadie et al., 2010 p. 494). Nonetheless, such an advantage is weakened by a lack of consensus on how (and what) covariates should be chosen. Due to this approach's relative infancy, there are not enough papers to formally test for specification searching (Brodeur et al., 2016). On the one hand, using all lagged outcome variables avoids the problem of omitting potentially irrelevant covariates because it eliminates all other predictors effects (Kaul et al., 2018) so that the synthetic counterfactual is created regardless of the other predictor's values. This specification is the one that minimizes the Root Mean Squared Prediction Error (RMSPE) in the pre-treatment period, and that is not subject to arbitrary decisions. On the other hand, it makes all the other covariates irrelevant, threatening the estimators' unbiasedness in the post-treatment predictions (Ferman et al., 2020). Given this lack of guidance, focusing on the specification that uses all the pre-treatment outcome lags as matching variables, is generally recommended (Ferman et al., 2020) unless there is a strong prior belief that it is crucial to balance on a specific set of covariates (as in the present context). Moreover, optimizing the dependent variable's pre-treatment fit and ignoring the covariates can be quite misleading: the more the covariates are truly influential for future values of the outcome, the larger a potential bias of the estimated treatment effect may become.\footnote{This result is essentially attributable to the fact that covariates are fitted rather poorly when all outcome lags are used, introducing a bias that can be substantial even for reasonably long-treatment time spans.} Therefore, economic theory and the researcher's intuition play a relevant part in the context of the ASCM as well. Building on the relevant literature on the topic discussed in the Introduction, we have considered the following sets of covariates: variables related to the resilience of the health care system (captured by \textit{hsp}), to the demographic structure of the population (represented by \textit{age} and \textit{hld}), as well as to the pandemic dynamics (epitomized by \textit{num} and \textit{mob}). Operatively, we first consider the specifications that differ only in the combinations of pre-treatment outcome values used as predictors. Specifically, we consider the following ones: (a0) all pre-treatment outcome values; (a1) odd pre-treatment outcome values only; (a2) even pre-treatment outcome values only; (a3) mean of all pre-treatment outcomes; (a4) the first half of the pre-treatment outcome values; (a5) the first three-fourths of the pre-treatment outcome values. We also consider a set of specifications from (b1) to (b5) where lagged dependent values as defined in (a1)-(a5) are augmented by structural time-invariant covariates (namely \textit{hsp}, \textit{age} and \textit{hld}). Finally, in the third group of specifications from (c1) to (c5), we extend the set of predictors by including lagged time-varying predictors averaged over time in a way which is consistent with the pre-treatment outcome values in (b1) to (b5), respectively. \\
As a preliminary step, we compute the Average Treatment Effect on the Treated (ATT) that is the average deviation of the counterfactual series from the actual one over the treatment period (from March 9 2020, to April 11 2020) for each specification (a0)-(c5) to identify the specifications with a negative and statistically significant gap in a way which is consistent with our priors. Since there is more than one possible specification that satisfies the conditions above, we follow the recommendation by Ferman et al. (2020) of presenting results for many different specifications, and in particular, we include the specification (a0) as a benchmark. The upper part of Table \ref{tab:tab1} presents an overview of all the ASCM specifications that we consider in the analysis, while the last row reports the associated p-value for the computed ATT for the corresponding specification.\footnote{In all the specifications, we select the hyperparameter $\lambda^{ridge}$ as the largest $\lambda$ within one standard error of the $\lambda$ that minimizes the cross-validation placebo fit CV($\lambda$) as discussed in Section \ref{SCM} above. The results obtained under the alternative rule of picking the minimal $\lambda$ are almost identical to those reported in the main text.}
\begin{center}
INSERT HERE TABLE \ref{tab:tab1} \\
\end{center}

\begin{figure}[ht]
\begin{center}
\includegraphics{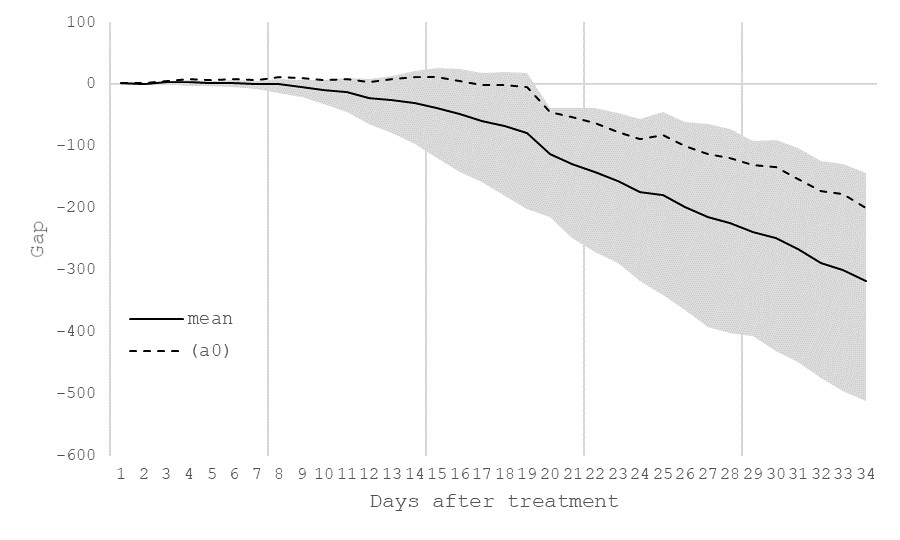}
\caption{\textbf{Gap plot for the benchmark case} \label{fig:fig1} \\
Note: The horizontal axis indicates the days after treatment. The dashed line is the difference between mortality rate in Italy and the synthetic control from the specification with all pre-treatment outcome values (a0). The solid black line is the gap plot obtained by averaging the synthetic control over all the alternative specifications. The shaded area is the confidence region computed as $\pm2.5$ times the
(median) absolute deviation from the median as suggested by Leys et al. (2013)}
\end{center}
\end{figure}

The results show that all of the ATTs are negative and statistically significant at the 5 percent level (or better), calling for a criterion to combine the test statistics for the individual specifications to distil them into a summary test statistic (Imbens and Rubin, 2015). Expressly, we assume that the test function is simply a weighted average of the test statistics for individual specifications. The same equally-weighted scheme is applied to combine each specification into a synthetic statistic (Christensen and Miguel, 2018; Cohen-Cole et al., 2009). In this vein, Figure \ref{fig:fig1} shows the treatment effects, defined as the differences between the mortality rate in Italy and the synthetic control over the evaluation period, averaged across all specifications (continuous black line), as well as the benchmark specification (a0) (dashed line). As expected, the mean value of the treatment effects suggests a strongly negative impact a few days after the intervention date. Moreover, there is preliminary evidence of statistical significance for these deviations from the actual path in the long-run according to the confidence region computed as $\pm2.5$ times the (median) absolute deviation from the median as suggested by Leys et al. (2013).
\begin{center}
INSERT HERE FIGURE \ref{fig:fig1} \\
\end{center}

\begin{figure}[ht]
\begin{center}
\includegraphics{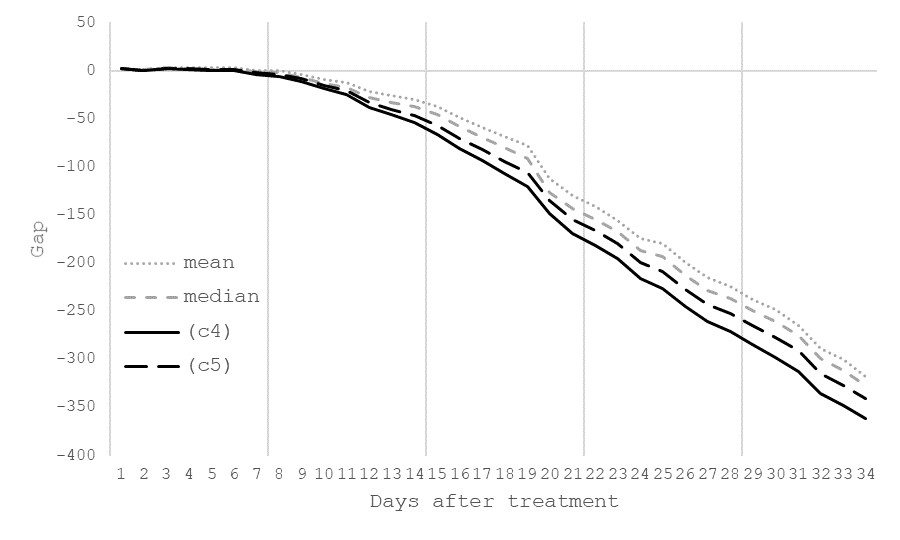}
\caption{\textbf{Gap plots for alternative specifications} \label{fig:fig2} \\
Note: The horizontal axis indicates the days after treatment. The solid black line is the difference between the mortality rate in Italy and the synthetic control (gap) from the specification (c4) with the first half of the pre-treatment outcome values averaged over time and structural time-invariant covariates (\textit{hsp}, \textit{age}, \textit{hld}), as defined in Section \ref{sec:specification}. The dotted and dashed lines represent the gap plot obtained by taking the average and the median of all the alternative specifications, respectively. The hyphenated line is the gap from the specification (c5) with the first three-fourths of the pre-treatment outcome values averaged over time and structural time-invariant covariates (\textit{hsp}, \textit{age}, \textit{hld}), as defined in Section \ref{sec:specification}.}
\end{center}
\end{figure}

While this finding gives indirect support to the effectiveness of non-pharmaceutical interventions in reducing the mortality rate, there is still a need for a criterion to select a given specification from a set of possible alternatives. It is well known that when covariates are expected to be useless in explaining the outcome, the recommended specification should use all pre-treatment outcome lags, i.e., specification (a0) (Kaul et al., 2018). However, if the control unit should also match several socio-economically relevant covariates, attention should be paid to the specifications allowing external predictors. Since b's are a particular case of their corresponding c's variant, we can safely restrict the focus on the latter group. A logical criterion to discriminate among the five remaining specifications with time-invariant and time-varying external predictors is given by each specification's lag structure. Since the pandemic dynamics are likely to affect the mortality rate with a temporal lag, specification (c3) is not the most desirable choice. Likewise, a lag structure based on either even or odd pre-treatment dates applied previously in SCM literature (see, for instance, Eren and Ozbeklik, 2016), is hard to rationalize in our context, suggesting that (c1) and (c2) are second-best options to alternative lag structures. This argumentation leads us to focus on the two remaining models: (c4) and (c5). In what follows, we pick (c4) as our preferred specification, while specification (c5) is used to check the robustness of our empirical findings. Accordingly, in our baseline model, the set of predictors includes time-invariant covariates, as well as the first half of the pre-treatment outcome values and time-varying covariates, averaged over the first half of the sample values, in a way similar to the empirical framework of reference in Cavallo et al. (2013). As Figure \ref{fig:fig2} shows, the gaps for our baseline specification closely resemble not only the one for specification (c5) but also the two summary statistics reported in Figure \ref{fig:fig1}: all in all, the temporal evolution of the treatment suggests a sort of delayed effect which becomes progressively negative as the days pass by.

\begin{center}
INSERT HERE FIGURE \ref{fig:fig2} \\
\end{center}

\subsection{Baseline specification: a primer}
\begin{figure}[ht]
\begin{center}
\includegraphics{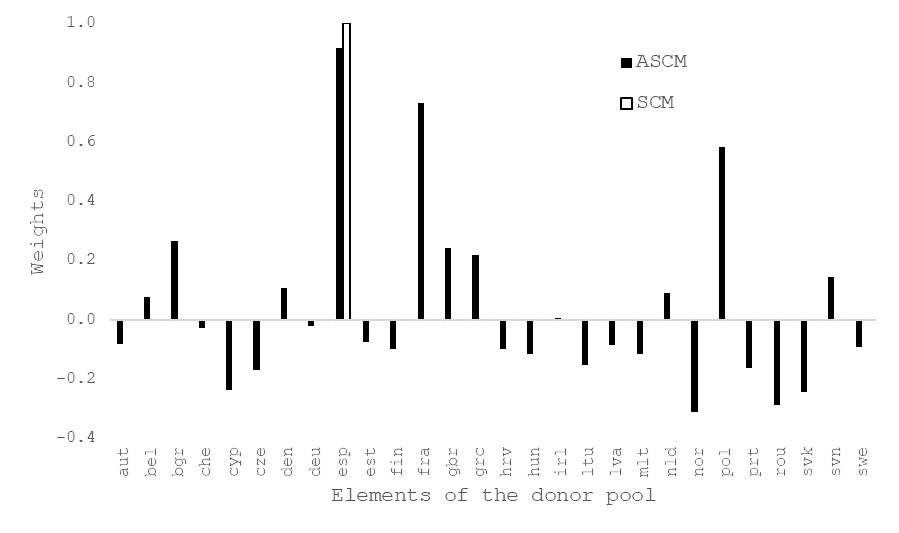}
\caption{\textbf{Comparison between SCM and ASCM weights} \label{fig:fig3} \\
Note: the picture reports the weights from the SCM (white) and from the ASCM (black)}
\end{center}
\end{figure}

Though suggestive, the visual evidence presented above is insufficient to ensure proper implementation of the ASCM. Its practical use calls for the fulfillment of three conditions. As for the first requirement, only the treated unit is affected by the policy change assessed over the post-treatment period (I); secondly, the counterfactual outcome can be approximated by a fixed combination of donor states (II); finally, the policy change has no effect before it is implemented (III).
 While the procedures to align country-specific variables to a common starting date as well as the definition of a general rule to identify the (average) treatment date for the donor set (i.e., the last observation of our sample) discussed in Section \ref{sec:datasource} above help to answer point (I), in what follows we focus on the remaining two conditions. As for point (II), Figure \ref{fig:fig3} reports the estimated weights according to both the ASCM and the canonical SCM for each country belonging to the donor pool. It emerges that the structure of the donor pool identified by the SCM consists of just one element (Spain) with weight zero attached to the remaining 27 countries. In contrast, in the case of the ASCM, a richer structure of the weights emerges. With regard to mortality rate this is in line with Cho (2020) who observes that death counts might be affected by measurement problems because some jurisdictions include both confirmed and probable cases of deaths, as opposed to others only reporting confirmed cases. In the specific case of Italy, Cerqua et al. (2020) claim that the official count of deaths due to COVID-19 is likely to under-report the phenomenon. In our context, the mechanics behind ASCM allows us to extrapolate-out of the convex hull and to assign non-zero weights to a higher number of countries. In more detail, there is confirmation of a relevant role for Spain (0.917), along with France (0.730) and Poland (0.583), with (relatively smaller) positive weights attached for several other countries as well (namely Belgium, Bulgaria, Denmark, Greece, the United Kingdom, the Netherlands, and Slovenia). In contrast, Germany, Switzerland, and Ireland do not contribute to the synthetic unit's construction, while the remaining 15 countries are associated with negative weights (ranging from -0.312 for Norway to -0.076 for Estonia). Overall, from an applied viewpoint, the ASCM can identify three European countries where the pandemic severity peaked compared to other donor pool candidate units. Simultaneously, resorting to the ASCM seems like a legit choice from a methodological standpoint due to the documented difficulty to build up a donor pool within the canonical SCM.
In such an occurrence, SCM and ASCM estimates tend to diverge. Specifically, ASCM estimates are likely to rely heavily on extrapolation from the convex hull of the control units in order to improve pre-treatment fit (Ben-Michael et al., 2020) by allowing for negative weights (if the treated unit is outside the convex hull) in place of the sparse and always non-negative weighting structure of the standard SCM. Moreover, resorting to a CV procedure makes it possible to obtain the optimal amount of extrapolation through the choice of $\lambda^{ridge}$ of condition (\ref{CV}).

\begin{center}
INSERT HERE FIGURE \ref{fig:fig3} \\
\end{center}
Turning to point (III), the synthetic outcome is expected to closely match the treated outcome's temporal profile during the pre-treatment period. Thus, as a preliminary step, Table \ref{tab:tab2} reports the average values over the pre-treatment period of the predictors for Italy (``actual'') and the counterfactual control (``synth''), where the latter is constructed with the ASCM weights assigned to the elements of the donor pool as detailed in Figure \ref{fig:fig3}. Overall, the synthetic control unit provides a much better-matched profile of Italy along the predictors compared to the simple average of all countries in the donor pool (donor), suggesting that the ASCM-based selection of weights is more appropriate as a control unit, rather than choosing subjectively the weights by means, for instance, the simple average across all the donor units.

							\begin{center}
				INSERT HERE TABLE \ref{tab:tab2} \\
							\end{center}

\begin{figure}[t!]
\centering
\begin{subfigure}[t]{0.5\textwidth}
\centering
\includegraphics[height=2.5in]{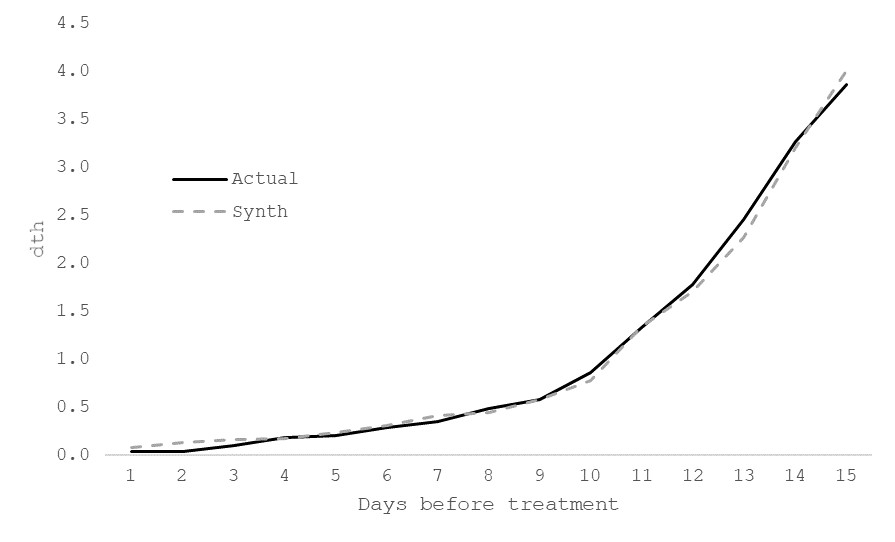}
\caption{Pre-treatment period} \label{fig:fig4a}
\end{subfigure}
\hfill
\begin{subfigure}[t]{0.5\textwidth}
\centering
\includegraphics[height=2.5in]{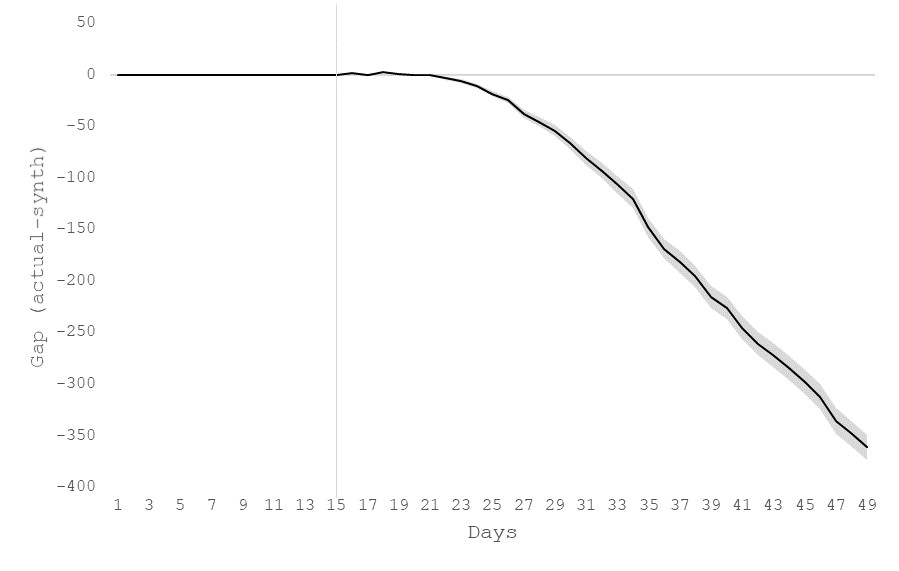}
\caption{Pre- and post-treatment periods} \label{fig:fig4b}
\end{subfigure}
 \caption{\textbf{Actual and synthetic Italy} \\ Note: In both panels, the horizontal axis indicates days after the death per million exceeds one. The profile of Italy is shown by the solid line while its synthetic counterfactual, the dashed line. The vertical line in Panel (\subref{fig:fig4b}) represents the lockdown date. The shaded area is the confidence interval computed by means of the jackknife+ approach of Barber et al. (2019).} \label{fig:fig4}
\end{figure}

Such a requirement is necessary to ensure that the comparison of the outcome paths during the post-treatment period provides insight into the effect of the treatments: when the dynamics of the synthetic control and the treated entity tend to diverge, then the treatment presumably caused the difference; in contrast, if both paths display similarities in the treatment period, the treatment does not appear to have affected the outcome. Figure \ref{fig:fig4} compares the temporal profile of mortality rate for both actual and synthetic Italy by disentangling pre- and post-treatment periods: the upper panel assesses the quality of fit in the pre-treatment period and refers to the first 15 daily observations since the case per million exceeds one (corresponding to the temporal window from February 23, 2020, to March 8, 2020); the lower panel displays the difference of the two series over the entire sample span which also includes the post-treatment period ranging from March 9, 2020 (when the policy intervention took place) to April 11, 2020.

\begin{center}
INSERT HERE FIGURE \ref{fig:fig4} \\
\end{center}

While the cumulative mortality rate in the synthetic control unit closely overlaps the actual series prior to the pre-treatment period, there is a visible divergence a few days after the policy intervention date (vertical line in the lower panel) when the synthetic unit starts following a much steeper path than the actual counterpart series. The resulting gap, defined as the difference between the actual series and its synthetic control, turns out to be negative and statistically significant with an ATT of -132.9 and a p-value of 0.000. In more detail, we find that the cumulative mortality rate in synthetic Italy exceeds 670 per million population roughly five weeks after the lockdown intervention, while the corresponding figure for actual Italy is as low as 310. This evidence suggests that with a (negative) gap of around 340 cases per million, Italy's mortality rate case would have been higher by over 115 percent had there not been the policy intervention. Moreover, the gap from the actual series turns out to be statistically significant at the 95 percent level according to the confidence interval computed by the jackknife+ approach of Barber et al. (2019).

\section{Sensitivity analysis}
\label{sec:sensitivity}
Confidence intervals as those reported in Figure \ref{fig:fig4} above have only recently been introduced (see also Cattaneo et al., 2019) in the SCM literature so that the statistical significance of the post-treatment gap is typically assessed by resorting to permutation techniques (the so-called placebo test; Abadie et al., 2010). Nonetheless, permutation-based tests can convey useful information in the present context to support our empirical findings. In what follows, we discuss three types of sensitivity tests: placebo in-space, placebo in-time, and leave-one-out tests. \\
Under the ``placebo in-space'' test, the ASCM is sequentially applied to each country in the donor pool as though it is a treated state, using the remaining members of the pool as before. The resulting placebo unit is thus compared with its synthetic counterpart. Comparing the difference between the treated unit and its synthetic control to the differences among placebo countries and their controls makes it possible to evaluate better the effectiveness  of policy intervention on the treated unit. \\ As the Root Mean Square Prediction Error (RMSPE) measures the gap between the variable of interest for the treated country and its synthetic counterpart, it is possible to calculate a set of RMSPE values for the pre-and post-treatment periods for each unit considered in the analysis.
Consequently, the RMSPE of the treated country after the treatment is expected to be large relative to its value before treatment. On the other hand, placebo units should not see a substantial increase in their RMSPE following the treatment. For this reason, Table \ref{tab:tab3}  reports the RMSPE pre/post-treatment ratio of each donor country divided by the same quantity computed for the treated country, Italy. Whenever the entry in the table is less than 1, it  indicates a relatively higher difficulty when forecasting future outcome values for Italy. The share of RMSPEs above one is then used to obtain a p-value for Italy, which measures the probability of observing a ratio as high as the one obtained for Italy if one were to pick a country at random from the potential controls.

\begin{center}
INSERT HERE TABLE \ref{tab:tab3} \\
\end{center}

\begin{figure}[ht]
\begin{center}
\includegraphics{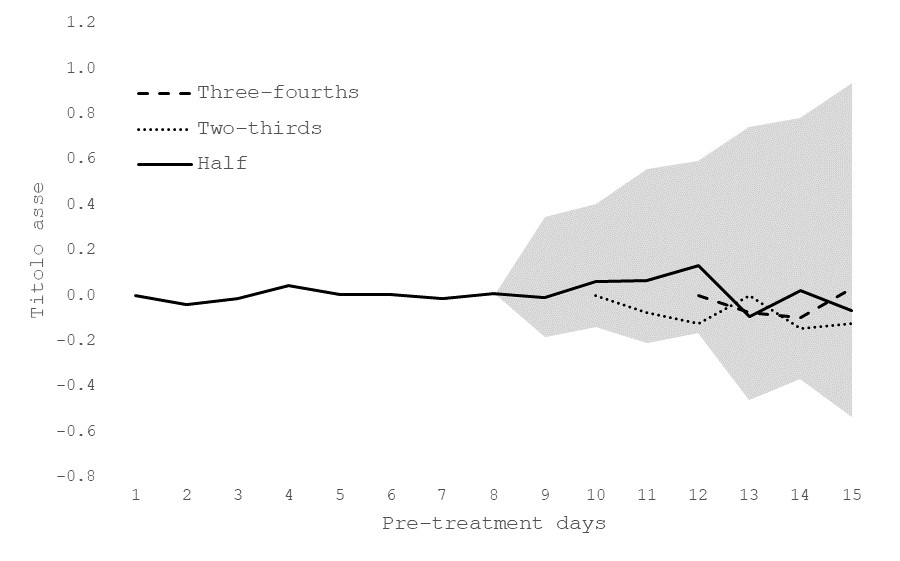}
\caption{\textbf{In-time placebo} \label{fig:fig5} \\
Note: The horizontal axis indicates the days after the death per million exceeds one. The black solid, dotted and dashed lines are the gap plot when the fictitious treatment date is assigned to half, two-thirds and three-fourths of the actual pre-treatment period, respectively. The shaded area is the 95 percent confidence interval computed by means of the jackknife+ approach of Barber et al. (2019).}
\end{center}
\end{figure}

Overall, the RMSPE ratios turn out to be well below the unit threshold, suggesting that the actual path of the treated states tends to diverge away from the synthetic control after the intervention in a much more substantial way than all countries belonging to the donor pool. The resulting p-value is (2/29=) 0.069 as it ranks second out of 29 countries, which falls within the conventional range of statistical significance used in the relevant literature. A remarkable exception is Belgium's case, with an RMSPE slightly above the unit; nonetheless, the associated ATT has an opposite sign to the expected one in a way similar to what emerges for the other eight countries. There is no evidence of a statistically significant ATT for three entities of the donor pool, while for the remaining cases, the estimated ATT ranges from -0.7 (for Croatia) to -34.5 (for Greece and Hungary). \\
As a further sensitivity test, we run the ``in-time placebo'' test, in which the donor pool remains fixed and the treated unit is always Italy, but the treatment date is re-assigned to occur during the pre-treatment period, as devised by Abadie et al. (2015). Moreover, this placebo model's sample period must end when the actual treatment occurred (day 15, in our context) to avoid capturing its effects. Operatively, the in-time placebo test is conducted under the assumption that the treatment occurred on day 8, roughly in the middle of our pre-treatment period. Apart from the lockdown date, we apply the baseline setup's exact setting to use the same predictor variables, including lagged outcome values for the first half of the pre-treatment period. As Figure \ref{fig:fig5} shows, our synthetic Italy for a placebo treatment on day 8 closely follows the path of actual Italy, not only during the first half of the baseline pre-treatment period but also in the second part of the sample with an estimated gap barely different from zero (continuous black line) also according to the 95 percent confidence region. Similar results are obtained when the fictitious treatment date is assigned to days 10 and 12 (corresponding to two-thirds and three-fourths of the baseline pre-treatment period, respectively). According to the dotted and dashed lines in Figure \ref{fig:fig5}, significant reductions in the mortality rate for these two fake lockdown dates cannot be found over the actual pre-treatment period. Overall, the in-time placebo test assures that the placebo estimate resembles the actual pre-treatment path closely enough to give us confidence that our main findings are not through chance, ruling out the possibility that the above-discussed difference between the synthetic and actual Italy arises for reasons other than the treatment.
\begin{center}
INSERT HERE FIGURE \ref{fig:fig5} \\
\end{center}

\begin{figure}[ht]
\begin{center}
\includegraphics{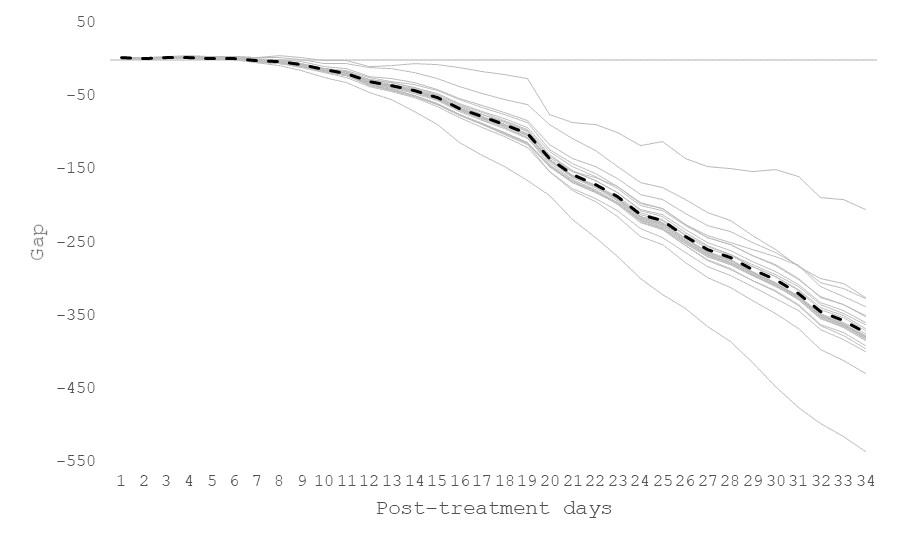}
\caption{\textbf{Leave-one-out} \label{fig:fig6} \\
Note: The horizontal axis indicates days after the death per million exceeds one. The black dashed line is the gap plot of the baseline specification obtained with the entire set of donor countries. The grey lines are the leave-one-out gap plots obtained by removing one country at a time from the donor pool of the baseline specification.}
\end{center}
\end{figure}

\begin{center}
INSERT HERE FIGURE \ref{fig:fig6} \\
\end{center}

The third sensitivity check we consider is the leave-one-out test (Abadie et al., 2015), where the model is iterated over to leave out one selected donor country each time to assess whether one of the donor units is driving the results. Figure \ref{fig:fig6} shows all leave-one-out synthetic gaps (thin grey lines) and the mean value across all of them (dashed line). It emerges that the average gap across all permutations closely matches the baseline gap that includes all donor states in terms of ATTs (-128.2 and -132.9, respectively), giving further support to the robustness of our findings.

\section{Concluding remarks and further research}
\label{sec:conclusion}
This paper is the first contribution that uses the ASCM to evaluate the effectiveness of non-pharmaceutical interventions against COVID-19. Evidence has been provided for Italy, the first Western country which has implemented shelter-in-place orders after China. The paper shows how the ASCM helps remove bias from a naive application of the canonical SCM. Indeed, the latter estimator shrinks the donor pool to only one country, i.e., Spain, generating \textit{de facto} an estimate of the effect as a bare difference in mean between Italy and Spain. Constraining SCM weights on the unit simplex may be too restrictive, especially when it is hard to reproduce accurate synthetic pre-treatment dynamics. Our empirical case falls precisely in this circumstance, and the ASCM overcomes the problem by assigning negative weights to some donor units. As already pointed out by Ben-Michael et al. (2020), since the ASCM removes the non-negativity constraint and allows for extrapolation outside of the convex hull, the pre-treatment fit from ASCM turns out to be at least as good as the pre-treatment fit from the SCM alone. Our evidence suggests that applied economists should compare weights obtained from the SCM and the ASCM and opt for the former only if the two are not entirely different. The socio-economic relevance of the issue analyzed in this contribution makes the importance of such a comparison even more evident because the estimates may significantly diverge, and different conclusions may be based on biased estimates. \\
The results report extensive evidence on the effectiveness of non-pharmaceutical interventions in avoiding human deaths and preventing health care systems from collapsing in the present COVID-19 era. According to our benchmark estimate, for each life lost, the policy has saved 1.15 lives. In other words, while the cumulative mortality rate has recorded 310 lives lost per million population, without lockdown policy, it would have been 670, that is 340 lives saved per million population. With 60 million as the 2020 Italian population in total, the policy has produced (60*340=) 20,400 lives saved. It is important to stress that this figure can be considered conservative because the sample span we can use for the econometric exercise is shorter than the total temporal horizon over which the policy has been implemented, due to the lack of non-treated units from a specific point in time onwards. Similar desirable results, albeit not comparable in magnitude, have been found by Friedson et al. (2020) for California, while for the case of Sweden, Cho (2020) estimates 25 percent of lives lost attributable to the non-treatment decision. Due to limited external validity and different methodologies applied, the figure by Cho (2020) is scantly comparable with ours; nevertheless, a more substantial effect in Italy is quite reasonable because of structural differences between the two countries in terms of (lower) endowment of hospital beds, (older) median age of the population and (larger) household average size (for Italy). As a possible extension, one could think of extending and projecting the findings up to the last day of the policy (i.e., other 36 extra days up to May 18, 2020) and re-calculate the total effect of the policy. One can also consider relating the findings of this paper to economic damages caused by the lockdown measure, while from a strictly methodological point of view, a further possible extension consists in constructing a formal test to test the equality in the mean of the weights generated by the SCM and the ASCM. These issues are beyond the scope of the present work and are left for further research.

\newpage

\begin{sidewaystable}
\caption{Full set of specifications
\\
\small
NOTE: \textit{dth(i)}: i-th lag of cumulative death counts per million population; \textit{hsp}: number of hospital beds per hundred thousand
population; \textit{age} median age; \textit{hld}: average household size; \textit{num(.)}: cumulative cases per million population; \textit{mob(.)}: mobility indicator; (.) indicates the average value over the lags chosen for the dependent variable: (*) stands for the average value over the entire pre-sample period; ATT= Average Treatment effect on the Treated.}
    \centering \label{tab:tab1}
		 \tiny{
    \begin{tabular}{|l|l|l|l|l|l|l|l|l|l|l|l|l|l|l|l|l|}
    \hline
		&\multicolumn{16}{c}{Specification}                            \\ \hline
         & (a0) & (a1) & (a2) & (a3) & (a4) & (a5) & (b1) & (b2) & (b3) & (b4) & (b5) & (c1) & (c2) & (c3) & (c4) & (c5) \\ \hline \hline
        dth(1) & * & * &  &  & * & * & * &  &  & * & * & * &  &  & * & * \\ \hline
        dth(2) & * &  & * &  & * & * &  & * &  & * & * &  & * &  & * & * \\ \hline
        dth(3) & * & * &  &  & * & * & * &  &  & * & * & * &  &  & * & * \\ \hline
        dth(4) & * &  & * &  & * & * &  & * &  & * & * &  & * &  & * & * \\ \hline
        dth(5) & * & * &  &  & * & * & * &  &  & * & * & * &  &  & * & * \\ \hline
        dth(6) & * &  & * &  & * & * &  & * &  & * & * &  & * &  & * & * \\ \hline
        dth(7) & * & * &  &  & * & * & * &  &  & * & * & * &  &  & * & * \\ \hline
        dth(8) & * &  & * &  &  & * &  & * &  &  & * &  & * &  &  & * \\ \hline
        dth(9) & * & * &  &  &  & * & * &  &  &  & * & * &  &  &  & * \\ \hline
        dth(10) & * &  & * &  &  & * &  & * &  &  & * &  & * &  &  & * \\ \hline
        dth(11) & * & * &  &  &  &  & * &  &  &  &  & * &  &  &  &  \\ \hline
        dth(12) & * &  & * &  &  &  &  & * &  &  &  &  & * &  &  &  \\ \hline
        dth(13) & * & * &  &  &  &  & * &  &  &  &  & * &  &  &  &  \\ \hline
        dth(14) & * &  & * &  &  &  &  & * &  &  &  &  & * &  &  &  \\ \hline
        dth(15) & * & * &  &  &  &  & * &  &  &  &  & * &  &  &  &  \\ \hline
        dth(*) &  &  &  & * &  &  &  &  & * &  &  &  &  & * &  &  \\ \hline
        hsp &  &  &  &  &  &  & * & * & * & * & * & * & * & * & * & * \\ \hline
        age &  &  &  &  &  &  & * & * & * & * & * & * & * & * & * & * \\ \hline
        hld &  &  &  &  &  &  & * & * & * & * & * & * & * & * & * & * \\ \hline
        num(.) &  &  &  &  &  &  &  &  &  &  &  & * & * & * & * & * \\ \hline
        mob(.) &  &  &  &  &  &  &  &  &  &  &  & * & * & * & * & * \\ \hline \hline
        ATT & -46.9 & -138.7 & -103.8 & -56.1 & -81.0 & -46.8 & -132.1 & -104.9 & -65.1 & -136.3 & -102.5 & -140.2 & -137.1 & -133.9 & -132.9 & -122.2 \\ \hline
        p-value & [0.000] & [0.000] & [0.000] & [0.000] & [0.000] & [0.000] & [0.000] & [0.000] & [0.000] & [0.000] & [0.000] & [0.000] & [0.000] & [0.000] & [0.000] & [0.000] \\ \hline
    \end{tabular}
					}
%
\end{sidewaystable}

\begin{table}
\caption{Balancing table.
\\
\small NOTE: \textit{dth(i)}: i-th lag of cumulative death counts per million population; \textit{hsp}: number of hospital beds per hundred thousand
population; \textit{age} median age; \textit{hld}: average household size; \textit{num(.)}: cumulative cases per million population; \textit{mob(.)}: mobility indicator.}
    \centering \label{tab:tab2}
    \begin{tabular}{|l|l|l|l|}
    \hline
         & Actual & Synth & Donor \\ \hline \hline
        dth(1) & 0.033 & 0.026 & 0.001 \\ \hline
        dth(2) & 0.033 & 0.079 & 0.008 \\ \hline
        dth(3) & 0.099 & 0.109 & 0.01 \\ \hline
        dth(4) & 0.182 & 0.124 & 0.011 \\ \hline
        dth(5) & 0.199 & 0.182 & 0.019 \\ \hline
        dth(6) & 0.282 & 0.253 & 0.023 \\ \hline
        dth(7) & 0.348 & 0.356 & 0.039 \\ \hline
        hsp & 3.180 & 3.641 & 4.777 \\ \hline
        age & 45.500 & 44.034 & 42.007 \\ \hline
        hld & 2.400 & 2.383 & 2.449 \\ \hline
        num(.) & 6.341 & 5.066 & 6.002 \\ \hline
        mob(.) & 4.143 & 4.943 & 2.492 \\ \hline
    \end{tabular}
\end{table}

\begin{table}
\caption{In-space placebo.
\\
\small Note: The column ``RMSPE ratio'' reports the post/pre-treatment RMSPE of each country of the donor pool relative to the same ratio for Italy. Whenever the entry is less than one, the relative difficulty in forecasting future outcome values after intervention for Italy is higher than the one for the country considered. ATT= Average Treatment effect on the Treated.}
    \centering \label{tab:tab3}
 \begin{tabular}{|l|l|l|l|}
    \hline
         &  RMSPE ratio &  ATT &  P-value \\ \hline \hline
        aut & 0.050 &  7.1 & [0.000] \\ \hline
        bel & 1.050 &  133.8 & [0.000] \\ \hline
        bgr & 0.006 &  -0.1 & [0.384] \\ \hline
        che & 0.102 &  14.1 & [0.000] \\ \hline
        cyp & 0.043 &  1.1 & [0.280] \\ \hline
        cze & 0.016 &  -1.8 & [0.000] \\ \hline
        den & 0.070 &  8.2 & [0.000] \\ \hline
        deu & 0.057 &  6.5 & [0.000] \\ \hline
        esp & 0.414 &  61.3 & [0.000] \\ \hline
        est & 0.014 &  -1.5 & [0.000] \\ \hline
        fin & 0.178 &  -20.6 & [0.000] \\ \hline
        fra & 0.067 &  3.3 & [0.107] \\ \hline
        gbr & 0.094 &  4.7 & [0.100] \\ \hline
        grc & 0.194 &  -25.0 & [0.000] \\ \hline
        hrv & 0.005 &  -0.7 & [0.000] \\ \hline
        hun & 0.244 &  -34.5 & [0.000] \\ \hline
        irl & 0.302 &  -42.2 & [0.000] \\ \hline
        ltu & 0.324 &  -43.0 & [0.000] \\ \hline
        lva & 0.054 &  -7.2 & [0.000] \\ \hline
        mlt & 0.013 &  -1.9 & [0.000] \\ \hline
        nld & 0.076 &  11.6 & [0.000] \\ \hline
        nor & 0.331 &  -45.8 & [0.000] \\ \hline
        pol & 0.890 &  115.9 & [0.000] \\ \hline
        prt & 0.041 &  -2.2 & [0.079] \\ \hline
        rou & 0.159 &  -23.0 & [0.000] \\ \hline
        svk & 0.071 &  10.9 & [0.000] \\ \hline
        svn & 0.013 &  -0.8 & [0.032] \\ \hline
        swe & 0.179 &  20.2 & [0.000] \\ \hline
    \end{tabular}
\end{table}
\end{document}